\catcode`\@=11

\font\teneufm=eufm10
\font\seveneufm=eufm7
\font\fiveeufm=eufm5
\newfam\eufmfam
\textfont\eufmfam=\teneufm  \scriptfont\eufmfam=\seveneufm
  \scriptscriptfont\eufmfam=\fiveeufm
\def\eufm@{\hexnumber@\eufmfam}

\font\tenmsa=msam10
\font\sevenmsa=msam7
\font\fivemsa=msam5
\font\tenmsb=msbm10
\font\sevenmsb=msbm7
\font\fivemsb=msbm5
\newfam\msafam
\newfam\msbfam
\textfont\msafam=\tenmsa  \scriptfont\msafam=\sevenmsa
  \scriptscriptfont\msafam=\fivemsa
\textfont\msbfam=\tenmsb  \scriptfont\msbfam=\sevenmsb
  \scriptscriptfont\msbfam=\fivemsb

\def\hexnumber@#1{\ifnum#1<10 \number#1\else
 \ifnum#1=10 A\else\ifnum#1=11 B\else\ifnum#1=12 C\else
 \ifnum#1=13 D\else\ifnum#1=14 E\else\ifnum#1=15 F\fi\fi\fi\fi\fi\fi\fi}

\def\msa@{\hexnumber@\msafam}
\def\msb@{\hexnumber@\msbfam}

\mathchardef\gx="2\eufm@78
\mathchardef\gg="2\eufm@67
\mathchardef\gm="2\eufm@6D

\mathchardef\gd="2\eufm@64

\mathchardef\boxdot="2\msa@00
\mathchardef\boxplus="2\msa@01
\mathchardef\boxtimes="2\msa@02
\mathchardef\square="0\msa@03
\mathchardef\blacksquare="0\msa@04
\mathchardef\centerdot="2\msa@05
\mathchardef\lozenge="0\msa@06
\mathchardef\blacklozenge="0\msa@07
\mathchardef\circlearrowright="3\msa@08
\mathchardef\circlearrowleft="3\msa@09
\mathchardef\rightleftharpoons="3\msa@0A
\mathchardef\leftrightharpoons="3\msa@0B
\mathchardef\boxminus="2\msa@0C
\mathchardef\Vdash="3\msa@0D
\mathchardef\Vvdash="3\msa@0E
\mathchardef\vDash="3\msa@0F
\mathchardef\twoheadrightarrow="3\msa@10
\mathchardef\twoheadleftarrow="3\msa@11
\mathchardef\leftleftarrows="3\msa@12
\mathchardef\rightrightarrows="3\msa@13
\mathchardef\upuparrows="3\msa@14
\mathchardef\downdownarrows="3\msa@15
\mathchardef\upharpoonright="3\msa@16

\mathchardef\downharpoonright="3\msa@17
\mathchardef\upharpoonleft="3\msa@18
\mathchardef\downharpoonleft="3\msa@19
\mathchardef\rightarrowtail="3\msa@1A
\mathchardef\leftarrowtail="3\msa@1B
\mathchardef\leftrightarrows="3\msa@1C
\mathchardef\rightleftarrows="3\msa@1D
\mathchardef\Lsh="3\msa@1E
\mathchardef\Rsh="3\msa@1F
\mathchardef\rightsquigarrow="3\msa@20
\mathchardef\leftrightsquigarrow="3\msa@21
\mathchardef\looparrowleft="3\msa@22
\mathchardef\looparrowright="3\msa@23
\mathchardef\circeq="3\msa@24
\mathchardef\succsim="3\msa@25
\mathchardef\gtrsim="3\msa@26
\mathchardef\gtrapprox="3\msa@27
\mathchardef\multimap="3\msa@28
\mathchardef\therefore="3\msa@29
\mathchardef\because="3\msa@2A
\mathchardef\doteqdot="3\msa@2B

\mathchardef\triangleq="3\msa@2C
\mathchardef\precsim="3\msa@2D
\mathchardef\lesssim="3\msa@2E
\mathchardef\lessapprox="3\msa@2F
\mathchardef\eqslantless="3\msa@30
\mathchardef\eqslantgtr="3\msa@31
\mathchardef\curlyeqprec="3\msa@32
\mathchardef\curlyeqsucc="3\msa@33
\mathchardef\preccurlyeq="3\msa@34
\mathchardef\leqq="3\msa@35
\mathchardef\leqslant="3\msa@36
\mathchardef\lessgtr="3\msa@37
\mathchardef\backprime="0\msa@38
\mathchardef\risingdotseq="3\msa@3A
\mathchardef\fallingdotseq="3\msa@3B
\mathchardef\succcurlyeq="3\msa@3C
\mathchardef\geqq="3\msa@3D
\mathchardef\geqslant="3\msa@3E
\mathchardef\gtrless="3\msa@3F
\mathchardef\sqsubset="3\msa@40
\mathchardef\sqsupset="3\msa@41
\mathchardef\trianglerighteq="3\msa@44
\mathchardef\trianglelefteq="3\msa@45
\mathchardef\bigstar="0\msa@46
\mathchardef\between="3\msa@47
\mathchardef\blacktriangledown="0\msa@48
\mathchardef\blacktriangleright="3\msa@49
\mathchardef\blacktriangleleft="3\msa@4A
\mathchardef\blacktriangle="0\msa@4E
\mathchardef\triangledown="0\msa@4F
\mathchardef\eqcirc="3\msa@50
\mathchardef\lesseqgtr="3\msa@51
\mathchardef\gtreqless="3\msa@52
\mathchardef\lesseqqgtr="3\msa@53
\mathchardef\gtreqqless="3\msa@54
\mathchardef\Rrightarrow="3\msa@56
\mathchardef\Lleftarrow="3\msa@57
\mathchardef\veebar="2\msa@59
\mathchardef\barwedge="2\msa@5A
\mathchardef\doublebarwedge="2\msa@5B
\mathchardef\angle="0\msa@5C
\mathchardef\measuredangle="0\msa@5D
\mathchardef\sphericalangle="0\msa@5E
\mathchardef\varpropto="3\msa@5F
\mathchardef\smallsmile="3\msa@60
\mathchardef\smallfrown="3\msa@61
\mathchardef\Subset="3\msa@62
\mathchardef\Supset="3\msa@63
\mathchardef\Cup="2\msa@64

\mathchardef\Cap="2\msa@65

\mathchardef\curlywedge="2\msa@66
\mathchardef\curlyvee="2\msa@67
\mathchardef\leftthreetimes="2\msa@68
\mathchardef\rightthreetimes="2\msa@69
\mathchardef\subseteqq="3\msa@6A
\mathchardef\supseteqq="3\msa@6B
\mathchardef\bumpeq="3\msa@6C
\mathchardef\Bumpeq="3\msa@6D
\mathchardef\lll="3\msa@6E

\mathchardef\ggg="3\msa@6F

\mathchardef\circledS="0\msa@73
\mathchardef\pitchfork="3\msa@74
\mathchardef\dotplus="2\msa@75
\mathchardef\backsim="3\msa@76
\mathchardef\backsimeq="3\msa@77
\mathchardef\complement="0\msa@7B
\mathchardef\intercal="2\msa@7C
\mathchardef\circledcirc="2\msa@7D
\mathchardef\circledast="2\msa@7E
\mathchardef\circleddash="2\msa@7F
\def\ulcorner{\delimiter"4\msa@70\msa@70 }
\def\urcorner{\delimiter"5\msa@71\msa@71 }
\def\llcorner{\delimiter"4\msa@78\msa@78 }
\def\lrcorner{\delimiter"5\msa@79\msa@79 }
\def\yen{\mathhexbox\msa@55 }
\def\checkmark{\mathhexbox\msa@58 }
\def\circledR{\mathhexbox\msa@72 }
\def\maltese{\mathhexbox\msa@7A }
\mathchardef\lvertneqq="3\msb@00
\mathchardef\gvertneqq="3\msb@01
\mathchardef\nleq="3\msb@02
\mathchardef\ngeq="3\msb@03
\mathchardef\nless="3\msb@04
\mathchardef\ngtr="3\msb@05
\mathchardef\nprec="3\msb@06
\mathchardef\nsucc="3\msb@07
\mathchardef\lneqq="3\msb@08
\mathchardef\gneqq="3\msb@09
\mathchardef\nleqslant="3\msb@0A
\mathchardef\ngeqslant="3\msb@0B
\mathchardef\lneq="3\msb@0C
\mathchardef\gneq="3\msb@0D
\mathchardef\npreceq="3\msb@0E
\mathchardef\nsucceq="3\msb@0F
\mathchardef\precnsim="3\msb@10
\mathchardef\succnsim="3\msb@11
\mathchardef\lnsim="3\msb@12
\mathchardef\gnsim="3\msb@13
\mathchardef\nleqq="3\msb@14
\mathchardef\ngeqq="3\msb@15
\mathchardef\precneqq="3\msb@16
\mathchardef\succneqq="3\msb@17
\mathchardef\precnapprox="3\msb@18
\mathchardef\succnapprox="3\msb@19
\mathchardef\lnapprox="3\msb@1A
\mathchardef\gnapprox="3\msb@1B
\mathchardef\nsim="3\msb@1C
\mathchardef\napprox="3\msb@1D
\mathchardef\nsubseteqq="3\msb@22
\mathchardef\nsupseteqq="3\msb@23
\mathchardef\subsetneqq="3\msb@24
\mathchardef\supsetneqq="3\msb@25
\mathchardef\subsetneq="3\msb@28
\mathchardef\supsetneq="3\msb@29
\mathchardef\nsubseteq="3\msb@2A
\mathchardef\nsupseteq="3\msb@2B
\mathchardef\nparallel="3\msb@2C
\mathchardef\nmid="3\msb@2D
\mathchardef\nshortmid="3\msb@2E
\mathchardef\nshortparallel="3\msb@2F
\mathchardef\nvdash="3\msb@30
\mathchardef\nVdash="3\msb@31
\mathchardef\nvDash="3\msb@32
\mathchardef\nVDash="3\msb@33
\mathchardef\ntrianglerighteq="3\msb@34
\mathchardef\ntrianglelefteq="3\msb@35
\mathchardef\ntriangleleft="3\msb@36
\mathchardef\ntriangleright="3\msb@37
\mathchardef\nleftarrow="3\msb@38
\mathchardef\nrightarrow="3\msb@39
\mathchardef\nLeftarrow="3\msb@3A
\mathchardef\nRightarrow="3\msb@3B
\mathchardef\nLeftrightarrow="3\msb@3C
\mathchardef\nleftrightarrow="3\msb@3D
\mathchardef\divideontimes="2\msb@3E
\mathchardef\varnothing="0\msb@3F
\mathchardef\nexists="0\msb@40
\mathchardef\mho="0\msb@66
\mathchardef\thorn="0\msb@67
\mathchardef\beth="0\msb@69
\mathchardef\gimel="0\msb@6A
\mathchardef\daleth="0\msb@6B
\mathchardef\lessdot="3\msb@6C
\mathchardef\gtrdot="3\msb@6D
\mathchardef\ltimes="2\msb@6E
\mathchardef\rtimes="2\msb@6F
\mathchardef\shortmid="3\msb@70
\mathchardef\shortparallel="3\msb@71
\mathchardef\smallsetminus="2\msb@72
\mathchardef\thicksim="3\msb@73
\mathchardef\thickapprox="3\msb@74
\mathchardef\approxeq="3\msb@75
\mathchardef\succapprox="3\msb@76
\mathchardef\precapprox="3\msb@77
\mathchardef\curvearrowleft="3\msb@78
\mathchardef\curvearrowright="3\msb@79
\mathchardef\digamma="0\msb@7A
\mathchardef\varkappa="0\msb@7B
\mathchardef\hslash="0\msb@7D
\mathchardef\hbar="0\msb@7E
\mathchardef\backepsilon="3\msb@7F
\def\Bbb{\ifmmode\let\next\Bbb@\else
 \def\next{\errmessage{Use \string\Bbb\space only in math mode}}\fi\next}
\def\Bbb@#1{{\Bbb@@{#1}}}
\def\Bbb@@#1{\fam\msbfam#1}

\catcode`\@=\active

\def\eps{{\epsilon}}

\def\<{\langle}
\def\>{\rangle}

\def\tens{\mathop{\otimes}}

\def\text#1{{\rm #1}}
\def\note#1{}

\documentstyle[11pt,epsf]{article}
\begin{document}

\newpage
\hfill Swan-Maths-00/2

\begin{center} {\Large THE BRAIDING FOR REPRESENTATIONS OF 
$q$-DEFORMED AFFINE $sl_2$.}
\\ \baselineskip 13pt{\ }
{\ }\\ E. J. Beggs \& P. R. Johnson\\
{\ } \\ Department of Mathematics\\
University of Wales, Swansea\\ Wales SA2 8PP
\end{center}

\begin{quote}%\baselineskip 13pt
\noindent{\bf ABSTRACT.} We compute the braiding for the `principal gradation'
of $U_q(\widehat{{\it sl}_2})$ for $|q|=1$ from first principles,
starting from the idea of a rigid braided tensor category. It is not necessary to
assume either the crossing or the unitarity condition, from S-matrix theory.
 We demonstrate the uniqueness of the normalisation of the braiding
under certain analyticity assumptions, and show that its
 convergence 
 is critically dependent on the number theoretic properties
of the number $\tau$ in the deformation parameter $q=e^{2\pi i\tau}$. 
We also examine the convergence using probability,
assuming a uniform distribution for $q$ on the unit circle.
\end{quote}

\section{Introduction}
The sine-Gordon S-matrix for the quantum scattering of solitons was solved
initially in the literature in \cite{ZZ}, by imposing  conditions on the
S-matrix called crossing and unitarity. These conditions were
previously observed in other S-matrices found using perturbative techniques
in quantum field theory \cite{Olive}, and were subsequently taken to be
axioms for the non-perturbative result. 

The bootstrap methods use the $U_q(\widehat{sl_2})$ Hopf algebra \cite{CP}, and
its spin half representation $W$, which is a space of functions taking values in
${\Bbb C}^2$, where the first component corresponds to `solitons' and the
second to `anti-solitons'.
Then $W\otimes W$ corresponds to a two soliton system, which can
interact by collision. There is an initial two soliton quantum state
in $W\otimes W$, and after the collision process we have a final quantum
state in $W\otimes W$. The scattering matrix gives a map 
$W\otimes W\rightarrow W\otimes W$, which sends the initial to the final
state. In terms of the Hopf algebra, this map is a braiding \cite{Ma:book}.

This scattering matrix is fairly easy to find up to a multiplication by a
scalar function, but the scalar function itself is  more difficult. We 
denote this scalar function by $a(z)$ where $z\in {\Bbb C}$. The crossing
condition in terms of this function becomes 
\begin{equation}
a(z)=a\Big(-{q\over z}\Big){(z-z^{-1})\over (zq^{-1}-z^{-1}q)} \label{star}
\end{equation}
and the unitarity condition is $a(z)a(z^{-1})=1$.

The crossing condition arises from physics by equating  a
scattering process with the same scattering process after 
rotating the space-time diagram of the collision by a right-angle. This rotation
involves a time reversal of one of the incoming and one of the outgoing
solitons, which implies that these are turned into anti-solitons. The
rotation of the diagram also implies that $z$, which corresponds roughly
to the relative momentum of the colliding  solitons (conserved in the
collision), is transformed to $-z^{-1}q$. The interested reader
can refer to the comprehensive book  \cite{Olive} for a complete
discussion. The article \cite{ZZ} is also an excellent review.

Zamolodchikov-Zamolodchikov\cite{ZZ} solved these equations (\ref{star})
to get  a formula for $a(z)$ in terms of a double infinite 
product of gamma functions. It was thought that this product
probably converged for all physical values of $z$, and all values
of $q$ where $|q|=1$. However no proof of this was provided. 
The convergence of this function has  also not been treated
in subsequent papers in the literature.

Here, in this paper, we find an alternative formula which is more
amenable to a convergence analysis. We find that the convergence
is highly delicate,  and the function converges with probability one,
for $q$ on the unit circle, i.e. $q=e^{2\pi i\tau}$, including convergence for all 
irrational algebraic values of $\tau$, and diverges for certain
transcendental values of $\tau$.

Another alternative formula for $a(z)$ was found in Johnson \cite{PRJ},
as an integral, or as a combination of `regularised' quantum dilogarithms.
This was done to make contact with semi-classical results for the scattering
which involve integrating classical time delays. However,
the convergence of this formula, or of the individual quantum dilogarithms,
 was also difficult to analyse because the contour integrals which one has
to do are difficult to perform,
involving sums over an infinite  double set of poles.
 
In this paper we shall consider the problem purely in terms of braidings
of the representations of a Hopf algebra, rather than invoking the
crossing and unitarity conditions of S-matrix theory. We begin with
the loop group of analytic functions used in the classical inverse scattering procedure
for sine-Gordon \cite{BJ}, and deform this by inclusion of a parameter $q$
\cite{Bernard}. 

\section{The universal enveloping algebra ${\cal H}$}
We begin with the loop group of analytic functions from
$\Bbb C^*$ to $SL_2(\Bbb C)$ which obey the symmetry condition 
$U\phi(-z)U^{-1}=\phi(z)$, where
\[
U\ =\ \left(\matrix{ 1 & 0\cr 0 & -1}\right)\ .
\]
The Lie algebra for this group has generators
$X_{\pm 1}$, $X_{\pm 2}$ and $H$, given by
\begin{eqnarray} 
X_{+1}(z)\,=\, \left( \matrix{ 0 & 0 \cr z & 0  } \right) &,&
X_{-1}(z)\,=\, \left( \matrix{ 0 & 1/z \cr 0 & 0  } \right)\quad\!,\quad \!
H(z)\,=\, \left(\matrix{ 1 & 0 \cr 0& -1}\right)\ ,  \cr
X_{+2}(z)\,=\, \left( \matrix{ 0 & z \cr 0 & 0  } \right) &,&
X_{-2}(z)\,=\, \left( \matrix{ 0 & 0 \cr 1/z & 0  } \right) \ .\label{gen}
\end{eqnarray} 
 These generators obey the usual
coproduct rule for the
universal enveloping algebra
of a  Lie algebra, namely $\Delta(X_{\pm 1})=X_{\pm 1}\tens 1+1\tens X_{\pm 1}$ etc. This rule
can be deformed by the inclusion of a parameter $q\in\Bbb C$ to give
\begin{eqnarray} 
\Delta H\ =\
 1\tens H+H\tens 1 &,& \Delta X_{\pm 1}\ =\ X_{\pm 1}\tens q^{-H/2}+q^{H/2}
\tens X_{\pm 1}\ ,\cr  \Delta X_{\pm 2}&=&X_{\pm 2}\tens q^{H/2}+q^{-H/2}
\tens X_{\pm 2}\ .\label{cop}
\end{eqnarray} 
The Lie algebra structure remains the same. 
If in addition we define a counit $\eps$ (which kills all the generators) and an
 antipode $S$ (which has $S(H)=-H$ and $S(X_{\pm n})=-q^{\pm 1}X_{\pm n}$) we can make
the universal enveloping algebra into a Hopf algebra,
which we denote ${\cal H}$. This is the so-called
`principal gradation' of  $U_q(\widehat{{\it sl}_2})$, which is actually
a subalgebra of  $U_q(\widehat{{\it sl}_2})$. We use the convention that $r=\sqrt q$ in the 
terms $q^{\pm H/2}$.

\section{Rigid braided tensor categories} Here we shall give a highly abbreveiated,
specialised and incomplete
account of rigid braided tensor categories, for a full account see \cite{Ma:book}. 

The representations of a Hopf algebra ${\cal H}$ form a category, with objects the 
representations, and the morphisms $\rho:V\to W$ are intertwining maps for 
the representations $V$ and $W$. This means that
$\rho$ is linear and that $\rho(h(v))=h(\rho(v))$
for all $h\in {\cal H}$. 

The tensor product
of two representations is also a representation. The action on $V\tens W$
is given by the coproduct $\Delta:{\cal H}\to{\cal H}\tens{\cal H}$. If we write
$\Delta h=\sum h_{(1)}\tens  h_{(2)}$, then $h(v\tens w)=\sum h_{(1)}(v)\tens  h_{(2)}(w)$.
The category contains an `identity object', the representation $\Bbb C$
with all generators having zero action. This means that 
category of representations forms a tensor or monoidal category.
(Technically we should also say that the associator is trivial.) 

If the category is {\it rigid}, for an object $V$ there is a dual object $V'$, and
an `evaluation' morphism ${\rm eval}:V'\tens V\to \Bbb C$
given by ${\rm eval}(\alpha,v)=\alpha(v)$. 

If the category is {\it braided}, for two objects $V$ and $W$ there is a morphism
$\Psi_{VW}:V\tens W\to W\tens V$. The braiding is functorial, which means that
if there is a morphism $\theta:V\to X$, then
the maps $(I\tens\theta)\Psi_{VW}:V\tens W\to W\tens X$
and $\Psi_{XW}(\theta\tens I):V\tens W\to W\tens X$ are the same. 

Figure 1 shows the standard diagramatic notation used for braided categories. 
Elements of representations are denoted by vertical lines. For
evaluation (a), note that $\Bbb C$ is traditionally denoted by an invisible line.
The braiding $\Psi_{VW}:V\tens W\to W\tens V$ is shown in (b),
and the rule for the functoriality of the braiding in (c).

\epsfbox{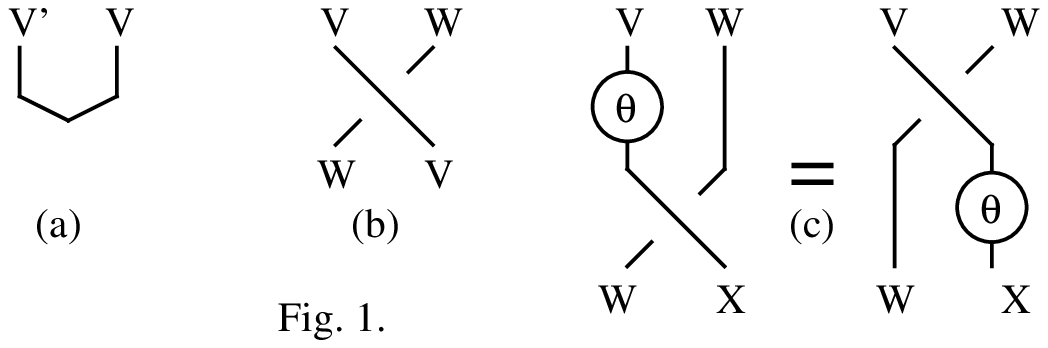}

The finite dimensional representations of a quasitriangular Hopf
algebra form a rigid braided tensor category. 
We shall assume that the representations of our Hopf algebra ${\cal H}$
also form a rigid braided tensor category, and this will allow us to 
explicitly calculate the braiding.

\section{The `standard' representation of ${\cal H}$}
Take $W$ to be a vector space of analytic functions $:\Bbb C^*\to \Bbb C^2$
(or at least analytic in a neighbourhood of zero and a neighbourhood of infinity),
which obeys the condition $Uw(-z)=w(z)$ for all $w\in W$. 
The algebra ${\cal H}$ acts on $W$ using matrix multiplication, $(hw)(z)=h(z)w(z)$,
for the five generators listed in (\ref{gen}).

We can consider a dual space $W'$ to $W$, which shall consist of analytic functions
$:\Bbb C^*\to \Bbb C^2$ which are defined for $|z|$ sufficiently small
and sufficiently large. 
Now define an evaluation map
${\rm eval}:W'\tens W\to \Bbb C$ by
\begin{eqnarray} 
{\rm eval}\Big(\left(\matrix{ f \cr g}\right)\tens \left(\matrix{ u \cr v}\right)\Big)
&=& \frac1{4\pi i} \oint_\gamma \Big( f(z)\, u(z)\,+\, g(z)\, v(z)\Big)\, \frac{dz}z\ ,\label{eval}
\end{eqnarray} 
where $\gamma$ consists of two  anticlockwise circular contours about $0$, one of large radius,
 and one of small radius. To find the action of ${\cal H}$ on $W'$ we use the fact that
the action commutes with $ {\rm eval}:W'\tens W\to \Bbb C$, and that the action of the generators is
zero on $\Bbb C$. This means that
\begin{eqnarray*} 
0 &=& H\Big({\rm eval} \Big(
\left(\matrix{ f \cr g}\right)\tens \left(\matrix{ u \cr v}\right)\Big)\Big) \cr
 &=&   {\rm eval} \Big(H\Big(
\left(\matrix{ f \cr g}\right)\tens \left(\matrix{ u \cr v}\right)\Big)\Big)\ =\ 
  {\rm eval} \Big( 
H\left(\matrix{ f \cr g}\right)\tens \left(\matrix{ u \cr v}\right)\,+\,
\left(\matrix{ f \cr g}\right)\tens H\left(\matrix{ u \cr v}\right)\Big)\ ,
\end{eqnarray*} 
so we get
\begin{eqnarray*} 
{\rm eval} \Big( 
H\left(\matrix{ f \cr g}\right)\tens \left(\matrix{ u \cr v}\right)\Big) &=&
-\, {\rm eval} \Big( 
\left(\matrix{ f \cr g}\right)\tens \left(\matrix{ u \cr -v}\right)\Big) \cr
&=&
-\, \frac1{4\pi i } \oint \Big( f(z)\, u(z)\,-\, g(z)\, v(z)\Big)\, \frac{dz}{z}\ ,
\end{eqnarray*} 
so we deduce that
\[
H\left(\matrix{ f \cr g}\right)(z)\ =\ \left(\matrix{ -1 & 0 \cr 0 & 1}\right)
\left(\matrix{ f(z) \cr g(z)}\right)\ .
\]
Now we continue with the generator $X_{+ 1}$,
\begin{eqnarray*} 
0 &=&  {\rm eval} \Big(X_{+ 1}\Big(
\left(\matrix{ f \cr g}\right)\tens \left(\matrix{ u \cr v}\right)\Big)\Big) \cr
&=&  {\rm eval} \Big( 
X_{+ 1}\left(\matrix{ f \cr g}\right)\tens \left(\matrix{ u/r \cr vr}\right)\,+\,
\left(\matrix{ f/r \cr gr}\right)\tens X_{+ 1}\left(\matrix{ u \cr v}\right)\Big)\ ,
\end{eqnarray*} 
from which we get
\[
{\rm eval} \Big( 
X_{+ 1}\left(\matrix{ f \cr g}\right)\tens \left(\matrix{ u/r \cr vr}\right)\Big) 
\ =\  -\, {\rm eval} \Big( 
\left(\matrix{ f/r \cr gr}\right)\tens \left(\matrix{ 0 \cr zu}\right)\Big) \ =\ 
 -\, \frac r{4\pi i } \oint z\, g(z)\, u(z)\, \frac{dz}{z}\ .
\]
From this, and the corresponding calculations for the other generators, we see that
the action on $W'$ is given by the matrix multiplication
\begin{eqnarray} 
\Big(h\left(\matrix{ f \cr g}\right)\Big)(z)&=& \tilde h(z)
\left(\matrix{ f(z) \cr g(z)}\right)\ ,\label{dual}
\end{eqnarray} 
where the matrices $\tilde h(z)$ are given by
\begin{eqnarray} 
\tilde X_{+1}(z)\,=\, -q\left( \matrix{ 0 & z \cr 0 & 0  } \right) &,&
\tilde X_{-1}(z)\,=\, -q^{-1}\left( \matrix{ 0 & 0 \cr 1/z & 0  } \right)\quad\!,\quad \!
\tilde H(z)\,=\, -\left(\matrix{ 1 & 0 \cr 0& -1}\right)\ ,  \cr
\tilde X_{+2}(z)\,=\, -q\left( \matrix{ 0 & 0 \cr z & 0  } \right) &,&
\tilde X_{-2}(z)\,=\, -q^{-1}\left( \matrix{ 0 & 1/z \cr 0 & 0  } \right) \ .\label{dualmat}
\end{eqnarray} 

There is a morphism $\theta:W\to W'$ given by
\begin{eqnarray} 
\theta(k)(z)&=& \left(\matrix{0& 1\cr 1& 0}\right)\, k(-zq)\ .\label{morph}
\end{eqnarray} 
To verify this we check that it commutes with the actions of the generators,
for example;
\begin{eqnarray*} 
(X_{+1}\theta(\left(\matrix{ u \cr v}\right)))(z) &=&
  -\, \left(\matrix{ 0 & qz \cr 0 & 0}\right)\, \theta(\left(\matrix{ u \cr v}\right))(z) 
\ =\ 
  -\, \left(\matrix{ 0 & qz \cr 0 & 0}\right)\, 
\left(\matrix{ v(-zq) \cr u(-zq)}\right)\ ,  \cr
\theta(X_{+1}\left(\matrix{ u \cr v}\right))(z)\ &=& \left(\matrix{0& 1\cr 1& 0}\right)\, 
(X_{+1}\left(\matrix{ u \cr v}\right))(-zq)\ =\   \left(\matrix{0& 1\cr 1& 0}\right)\, 
 \left(\matrix{0& 0\cr -q z & 0}\right)\, \left(\matrix{ u(-zq) \cr v(-zq)}\right) \ ,
\end{eqnarray*} 
which are equal as required.

\section{The braidings} We use the convention for tensor products that 
$\Bbb C^2\tens \Bbb C^2\cong \Bbb C^4$ and
$M_2\tens M_2\cong M_4$, where
\begin{eqnarray} 
\left(\matrix{u \cr v}\right) \tens \left(\matrix{u' \cr v'}\right) 
\cong \left(\matrix{uu' \cr uv' \cr vu' \cr vv'}\right)\ ,\quad
\left(\matrix{a & b \cr c & d}\right) \tens \left(\matrix{a' & b' \cr c' & d'}\right) 
\cong \left(\matrix{aa' & a b' & b a' & b b' \cr ac' & a d' & b c' & b d'
 \cr ca' & c b' & d a' & d b' \cr cc' & c d' & d c' & d d' }\right)
\ .\label{tens}
\end{eqnarray} 
Then we can consider $W\tens W$ or $W'\tens W$ as a space of analytic maps from a subset of
$\Bbb C^*\times\Bbb C^*$ to $\Bbb C^4$. 
Now if $k\in W'\tens W$ we have
\[
{\rm eval}(k)\ =\ \frac1{4\pi i} \oint \left(\matrix{ 1 & 0 & 0 & 1}\right)\,
k(x,x)\,\frac{dx}{x}\ .
\]
By the coproduct rule $H$ and (for example) $X_{+1}$ act on $W\tens W$ by
matrix multiplication
\begin{eqnarray*} 
(Hk)(x,y) &=& (H(x)\tens I_2+I_2\tens H(y))k(x,y)\ ,  \\
(X_{+1}k)(x,y) &=& (X_{+1}(x)\tens q^{-H(y)/2}+q^{H(x)/2}\tens X_{+1}(y))\, k(x,y)\ ,
\end{eqnarray*} 
and they act on $W'\tens W$ by 
\begin{eqnarray*} 
(Hk)(x,y) &=& (\tilde H(x)\tens I_2+I_2\tens H(y))k(x,y)\ ,  \\
(X_{+1}k)(x,y) &=& (\tilde X_{+1}(x)\tens q^{-H(y)/2}+q^{\tilde H(x)/2}\tens X_{+1}(y))\, k(x,y)\ .
\end{eqnarray*}

 The braiding $\Psi_{WW}:W\tens W\to W\tens W$
will be assumed to have the form $(\Psi_{WW}k)(x,y)=M(x,y)k(y,x)$, where
$M(x,y)$ is a $4\times 4$ matrix. Since the braiding is a morphism
we must have $\Psi_{WW}(hk)=h(\Psi_{WW}k)$ for the five generators $h$ and all $k\in  W\tens W$. 
The cases for $h=H$ and $X_{+1}$ are given below:
\begin{eqnarray*} 
M(x,y)\,  ( H(y)\tens I_2+I_2\tens H(x)) &=&  ( H(x)\tens I_2+I_2\tens H(y))\, M(x,y)\ , \\
M(x,y)\, ( X_{+1}(y)\tens q^{-H(x)/2}+q^{H(y)/2}\tens X_{+1}(x)) &=&
 ( X_{+1}(x)\tens q^{-H(y)/2}+q^{ H(x)/2}\tens X_{+1}(y))\, M(x,y).
\end{eqnarray*} 
A simple calculation will show that these five conditions determine the matrix
 $M(x,y)$ up to a complex
multiple, and we find
\[
(\Psi_{WW}k)(x,y)\ =\ a(x,y)\, \left( \matrix {
1 & 0  & 0  & 0 \cr 0 & 
\frac{xy(q^2-1)}{q^2x^2-y^2} &
\frac{q(x^2-y^2)}{q^2x^2-y^2} & 0 \cr
0 & \frac{q(x^2-y^2)}{q^2x^2-y^2} &
\frac{xy(q^2-1)}{q^2x^2-y^2} & 0 \cr
0 & 0 & 0 & 1}\right)\, k(y,x)\ ,
\]
where $a(x,y)$ is complex valued.
Note that $(\Psi_{WW}^2k)(x,y)=a(x,y)a(y,x)k(x,y)$. 
In the same manner we can determine
the braiding $\Psi_{W'W}:W'\tens W\to W\tens W'$  to be
$(\Psi_{W'W}k)(x,y)=N(x,y)k(y,x)$, where the matrix $N(x,y)$ is given by
\[
(\Psi_{W'W}k)(x,y)\ =\ \frac{c(x,y)}{q^2 y^2-x^2}\, \left( \matrix {
q^2 y^2-x^2 & 0  & 0  & (q^2-1)xy \cr 0 & 
0 &
q(y^2-x^2) & 0 \cr
0 & q(y^2-x^2) &
0 & 0 \cr
(q^2-1)xy & 0 & 0 & q^2 y^2-x^2}\right)\, k(y,x)\ ,
\]
where $c(x,y)$ is another complex valued function. 

Now we use the fact that as the braiding is functorial, it must commute
with the evaluation morphism. We see that the maps
$(I\tens{\rm eval})(\Psi_{W'W}\tens I)(I\tens \Psi_{WW})$
and ${\rm eval}\tens I:W'\tens W\tens W\to W$ are the same. 
In terms of the standard pictures, this is fig.\ 2.

\epsfbox{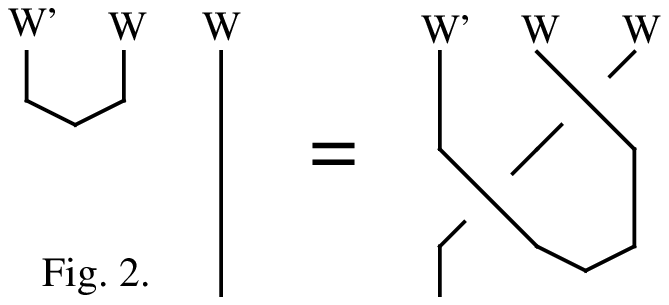}

Now,
identifying $W'\tens W\tens W$ with maps from subsets of $(\Bbb C^*)^3$ to
$\Bbb C^2\tens \Bbb C^2\tens \Bbb C^2$, we get
\begin{eqnarray} 
((\Psi_{W'W}\tens I)(I\tens \Psi_{WW})k)(x,y,z) &=&
(N(x,y)\tens I_2)((I\tens \Psi_{WW})k)(y,x,z)  \cr
&=& (N(x,y)\tens I_2)\, (I_2\tens M(x,z))\, k(y,z,x)\ ,
\end{eqnarray} 
and applying $I\tens{\rm eval}$ to this gives
\[
 \frac1{4\pi i} \oint (I_2\tens\left(\matrix{ 1 & 0 & 0 & 1}\right))\,
 (N(x,z)\tens I_2)\, (I_2\tens M(x,z))\, k(z,z,x)\,\frac{dz}{z}\ ,
\]
and some matrix multiplication shows that this is
\[
 \frac1{4\pi i} \oint a(x,z)\, c(x,z)\, (\left(\matrix{ 1 & 0 & 0 & 1}\right)\tens I_2)\,
k(z,z,x)\, \frac{dz}{z}\ .
\]
Just applying ${\rm eval}\tens I$ to $k$ gives
\[
 \frac1{4\pi i} \oint (\left(\matrix{ 1 & 0 & 0 & 1}\right)\tens I_2)\,
k(z,z,x)\, \frac{dz}{z}\ ,
\]
and since these must be the same for all choices of $k$ we deduce that 
$c(x,z)=1/a(x,z)$. 

Now use the fact that the braiding commutes with the morphism
$\theta:W\to W'$ in (\ref{morph}), so $(I\tens\theta)\Psi_{WW}$ and
$\Psi_{W'W}(\theta\tens I):W\tens W\to W\tens W'$ are the same. Then
\begin{eqnarray} 
((I\tens\theta)\Psi_{WW}k)(x,y) &=& (I_2\tens \left(\matrix{ 0 & 1 \cr 1 & 0 }\right))
(\Psi_{WW}k)(x,-qy) \cr 
&=& (I_2\tens \left(\matrix{ 0 & 1 \cr 1 & 0 }\right))\, M(x,-qy)\, k(-qy,x)\ , \cr
(\Psi_{W'W}(\theta\tens I)k)(x,y) &=& N(x,y)\, ((\theta\tens I)k)(y,x) \ ,\cr
&=& N(x,y)\, (\left(\matrix{ 0 & 1 \cr 1 & 0 }\right)\tens I_2)\, k(-qy,x)\ .
\end{eqnarray} 
From some more matrix multiplication, this is true if 
\begin{eqnarray} 
\frac{1}{a(x,y)} &=& a(x,-qy)\, \frac{x^2-q^2y^2}{q(x^2-y^2)}\ .\label{def}
\end{eqnarray}

\section{The normalisation of the braiding, $|q|=1$.}

In this section we find the normalisation $a(x,y)$. 
For the moment we shall suppose that the value of $x$ is fixed. 
 Suppose that a solution
$a(x,y)$ of (\ref{def}) is an analytic function of $y$ (except for isolated singularities)
 in some annulus centered on zero
contained in the region $|y|>|x|$. We can narrow the annulus down until it no longer contains any
isolated singularities or zeros. For convenience we set $z=x/y$
(so $|z|<1$), and $f(z)=a(x,x/z)$.
Then $f(z)$ satisfies the equation
\begin{eqnarray} 
f(z)f(-z/q) &=& q\, \frac{z^2-1}{z^2-q^2}\ .\label{def2}
\end{eqnarray} 
Now $f(z)$ will have a winding number $\omega\in\Bbb Z$ around zero
as $z$ winds once around zero. The function $c(z)=z^{-\omega}f(z)$ will have zero
winding number,
so its log, $b(z)=\log(c(z))$, will be analytic (and single valued) in the annulus. Now $c(z)$
obeys the equation
$c(z)c(-z/q)=q(-z^2/q)^{-\omega}(z^2-1)/(z^2-q^2)$, so we must have
\[
b(z)\,+\, b(-z/q)\ =\ \log(1-z^2)\,-\, \log(1-z^2/q^2)\,-\,\log(q)\,-\,\omega\log(-z^2/q)\ ,
\]
as $-z/q$ is in the annulus if $z$ is, because $|q|=1$. 
Since all the other functions are single valued in the annulus, we must have $\omega=0$. 
Now we can take the Laurent expansion of both sides in the annulus,
and compare coefficients, to get the {\it unique} solution
(up to the addition of a multiple of $\pi i$)
\begin{eqnarray} 
b(z)\ =\ -\frac12\log(q)\,+\,\sum_{n>0} \frac1n\, \frac {1-q^{2n}}{1+q^{2n}}\, z^{2n}\ ,\quad
|z|<1\ ,\  z\in{\rm  annulus}\ .\label{equplus}
\end{eqnarray} 
By the uniqueness of analytic continuation, the exponential of this formula must coincide with
$f(z)$ in a disk from zero up to the radius of convergence of the series. We also conclude that
$f(z)$ did in fact not have any zeros or isolated singularities in this disk. 

In the same manner, if
$f(z)$ were analytic (except for isolated singularities) in some annulus centered on zero
outside the unit disk, we could conclude that on that annulus
(again up to the addition of a multiple of $\pi i$)
\begin{eqnarray} 
b(z)\ =\ \frac12\log(q)\,-\,\sum_{n>0} \frac1n\, \frac {1-q^{2n}}{1+q^{2n}}\, z^{-2n}\ ,\quad
|z|>1\ .\label{equminus}
\end{eqnarray} 
We can now see
explicitly from the series (on the assumption that the series converge anywhere) that
$b(z)=-b(1/z)$ plus a multiple of $\pi i$, i.e.\ that $1/f(z)=\pm f(1/z)$. 

Now we can look at different values of $x$. 
As a consequence of the analyticity condition we assumed for $a(x,y)$,
it can be seen that $a(x,y)$ in fact only depends on $z=x/y$. We shall abuse our previous 
notation by referring to $a(z)$.

\section{The convergence of the series by number theory, $|q|=1$.}
The radius of convergence $R$ of the series (\ref{equplus}) is given by
\begin{eqnarray} 
R^{-1}\ =\ \limsup_{n\to+\infty} \left| \frac1n\,\cdot\, \frac {1-q^{2n}}{1+q^{2n}} \right|^{1/(2n)}
\ =\ \limsup_{n\to+\infty} \left|  \frac {1-q^{2n}}{1+q^{2n}} \right|^{1/(2n)}\ .\label{rad}
\end{eqnarray} 
We see that the series is not even defined if $q$ is an even root of $-1$. For any other root of unity
the series has radius of convergence $1$,
except for $q=\pm 1$, when the series terminates. Now consider the case
$q=e^{2\pi i\tau}$, where $\tau$ is irrational. 

Call $n\in\Bbb N$ type 1 if $|1-q^{2n}|<1$, which implies that $|1+q^{2n}|\ge 1$. Then
\[
\limsup_{n\to+\infty,\, {\rm type\, 1}} \left|  \frac {1-q^{2n}}{1+q^{2n}} \right|^{1/(2n)}
\ \le\ \limsup_{n\to+\infty,\, {\rm type\, 1}} 1\ =\ 1\ .
\]
Call $n\in\Bbb N$ type 2 if $|1-q^{2n}|\ge 1$, in which case
\[
\limsup_{n\to+\infty,\, {\rm type\, 2}} \left|  \frac {1-q^{2n}}{1+q^{2n}} \right|^{1/(2n)}
\ \ge\ \limsup_{n\to+\infty,\, {\rm type\, 2}}  \left|  \frac {1}{1+q^{2n}} \right|^{1/(2n)}
\ \ge\ 1\ ,
\]
and we deduce that 
\[
R^{-1}\ =\ \limsup_{n\to+\infty,\, {\rm type\, 2}} \left|  \frac {1-q^{2n}}{1+q^{2n}} \right|^{1/(2n)}\ .
\]
Then we find
\[
\limsup_{n\to+\infty,\, {\rm type\, 2}} \left|  \frac {1}{1+q^{2n}} \right|^{1/(2n)}
\ \le\ R^{-1}\ \le\ 
\limsup_{n\to+\infty,\, {\rm type\, 2}} \left|  \frac {2}{1+q^{2n}} \right|^{1/(2n)}\ ,
\]
which implies
\begin{eqnarray} 
R^{-1}\ =\ \limsup_{n\to+\infty,\, {\rm type\, 2}} \left|  \frac {1}{1+q^{2n}} \right|^{1/(2n)}
\ =\  \limsup_{n\to+\infty} \left|  \frac {1}{1+q^{2n}} \right|^{1/(2n)}\ .\label{roc}
\end{eqnarray} 
 If we let $d(\tau,n)$ be the minimum distance from $4\pi i n\tau$
to an odd multiple of $\pi i$, then
$d(\tau,n)\ge |1+q^{2n}|\ge \frac2\pi d(\tau,n)$, so
\begin{eqnarray} 
R\ =\ \liminf_{n\to+\infty} \left( \min_{p\ {\rm odd}}|4\pi i n\tau-p\pi i| \right)^{1/(2n)}
\ =\ \liminf_{n\to+\infty} \left( \min_{p\ {\rm odd}}|\tau-\frac{p}{4n}| \right)^{1/(2n)}\ \le\ 1\ .
\label{r2}
\end{eqnarray} 
We see that the radius of convergence of the power series is dependent on how well
$\tau$ can be approximated by rational numbers. Fortunately many results are
known in this area \cite{HW}, and we shall use one of these now.

Suppose that the irrational number $\tau\in\Bbb R$ is {\it algebraic} of degree $k>1$ (this means that
it is a root of a polynomial of degree $k$ with integer coefficients). 
Then there is a constant $K>0$
so that for all $n$ and all $p\in\Bbb Z$,
\[
|\tau-\frac{p}{4n}|\ \ge\ \frac K{(4n)^k}\ .
\]
From the formula for $R$ above (\ref{r2}), 
\[
R\ \ge\ \liminf_{n\to+\infty} \left( \frac K{2^k\ (2n)^k} \right)^{1/(2n)}\ =\ 1\ ,
\]
so we conclude that for any irrational algebraic number $\tau$, the  radius of convergence is 1. 

To get a radius of convergence less than 1, we shall
have to create an irrational  number with very good rational approximations.
Let
\begin{eqnarray} 
\tau\ =\ \sum_{s\ge 1} \frac 1{4m_s}\ ,
\label{taudef}
\end{eqnarray} 
where the strictly positive integers $m_s$
have the property that $4 m_s$ divides $m_{s+1}$ for all $s\ge 1$. 
If we set $n=m_t$, then
\[
\frac n{m_{t+1}}\ \le\ |l-4n\tau|\ \le\ \frac{2n}{m_{t+1}}
\]
for $l=\sum_{t\ge s\ge 1} m_t/m_s$ an odd integer, so 
\[
\frac1{4m_{t+1}}\ \le\ \min_{p\ {\rm odd}}|\tau-\frac{p}{4n}|\ \le\ \frac1{2m_{t+1}}\ .
\]
Then by (\ref{r2}),
\[
R\ \le\ \liminf_{t\to\infty}\left( \frac1{m_{t+1}} \right)^{1/(2m_t)}\ .
\]
If we set $m_1=1$ and $m_{s+1}=2^{2sm_s}$
for all $s\ge 1$, then the  radius
of convergence  of the series (\ref{equplus})
 for $\tau$ given by (\ref{taudef}) is zero. Also $\tau$ is too closely approximated
by non-equal rational numbers to be rational itself.

\section{The convergence of the series by probability, $|q|=1$.}

We consider the probability that the radius of convergence $R$
of the series (\ref{equplus})
is 1, given that $q$ has a uniform distribution on the 
circle (equivalently, $\tau$ has a uniform distribution
on $[0,1]$). From (\ref{r2}),  for $0<s<1$:
\begin{eqnarray} 
P[R>s]\ =\ P[\liminf_{n\to\infty} \min_{{p\ {\rm odd}}} |\tau-
\frac p{4n}|^{1/(2n)} > s]
\ =\ \lim_{m\to\infty} P[\inf_{n\ge m}\, 
\min_{{p\ {\rm odd}}} |\tau-
\frac p{4n}|^{1/(2n)} > s]\ .\label{infs}
\end{eqnarray} 
For any random variable $X_n$ and $s<t<1$,
\[
(\forall n\ge m\quad X_n > t)  \ \Rightarrow \ 
\inf_{n\ge m} X_n >s\ ,
\]
or alternatively
\[
P[X_n>t\quad \forall n\ge m]\ \le\ P[\inf_{n\ge m} X_n >s]\ .
\]
Then from (\ref{infs}),
\[
P[R> s]\ \ge\ \lim_{m\to\infty} P[ \min_{{p\ {\rm odd}}} |\tau-
\frac p{4n}|^{1/(2n)} > t \quad\forall n\ge m]\ .
\]
If we define
\[
A_n(x)\ =\ \{\tau\in[0,1]:\min_{{p\ {\rm odd}}} |\tau-
\frac p{4n}|\le x\}\ ,
\]
then (with superscript $c$ denoting complement)
\[
P[R>s]\ \ge\  \lim_{m\to\infty} P[\cap_{n\ge m}
A_n(t^{2n})^c]\ .
\]
Then by taking complements
\[
P[R\le s]\ \le\ \lim_{m\to\infty} P[\cup_{n\ge m} A_n(t^{2n})]
\ \le\  \lim_{m\to\infty} \sum_{n\ge m} P[A_n(t^{2n})]\ ,
\]
so if the series
\begin{eqnarray} 
\sum_{n\ge 1} P[A_n(t^{2n})] \label{ser}
\end{eqnarray} 
converges, then $R>s$ with probability one. If we write
\[
A_n(x)\ =\ \{\tau\in[0,1]:\min_{{p\ {\rm odd}}} |4n\tau-
p|\le 4nx\}\ ,
\]
then, if $4nx\le 1$, in the interval $[0,4n]$ there are $2n$
odd integers $p$, and each has a disjoint interval of length
$8nx$ about it satisfying the inequality above.
From this we find $P[A_n(x)]=4nx$, so the sum (\ref{ser}) becomes
(with the exception of a finite number of terms at the beginning),
\[
\sum_{n\ge 1} 4n\, t^{2n}\ ,
\]
which
converges since $|t|<1$. We conclude that $P[R>s]=1$, and so $R=1$ with probability 
one. Since the algebraic numbers have measure zero, the series must have $R=1$ for 
many transcendental (non-algebraic) $\tau$. 

\section{The case $|q|\neq 1$.}
In this case we would get the same unique series {\it if}
the annulus in which $a(z)$ was analytic and free of zeros or singularities
was sufficiently wide. We would need to have both $z$ and $-z/q$ in the same annulus, so the ratio
of the outer and inner radii of the annulus would have to be greater than the larger of
$|q|$ and $1/|q|$. If this condition was satisfied, we would have the solutions
(\ref{equplus}) or (\ref{equminus}). However there may be other solutions
to the normalisation which would always have zeros or singularities in such wide annuli.

Let us examine the series (\ref{equplus}) for $|q|\neq 1$. Here
$\lim_{n\to+\infty}|(1-q^{2n})/(1+q^{2n})|=1$, so the series has radius of convergence 1. 
However now we can make an analytic continuation of the series. In the case $|q|<1$
we write, for any integer $k\ge 1$,
\begin{eqnarray} 
\frac{1-q^{2n}}{1+q^{2n}}\ =\ 1\,+\,2\sum_{m=1}^k (-1)^m\, q^{2mn}\,-\, \frac{2\,(-1)^k\,
q^{2(k+1)n}}{1+q^{2n}}\ ,\label{trans}
\end{eqnarray} 
and substituting this into (\ref{equplus}) gives
\[
b(z)\ =\ -\frac12\log(q)\,-\,\log(1-z^2)\,-\,2\sum_{m=1}^k (-1)^m\, \log(1-z^2q^{2m})
\,-\, \sum_{n\ge 1}\frac{2\,(-1)^k\,
q^{2(k+1)n}\, z^{2n}}{n\,(1+q^{2n})}\ .
\]
The last term in the formula tends to zero uniformly on any bounded set as $k\to\infty$, allowing
us to take the limit as $k\to\infty$ of the other terms. We then get the infinite product expansion,
valid everywhere in $\Bbb C^*$,
\begin{eqnarray} 
a_+(z) &=& \frac1{r}\cdot\frac1{(1-z^2)}\cdot\prod_{{\rm odd}\ m\ge 1}
\frac{(1-z^2q^{2m})^2}{(1-z^2q^{2(m+1)})^2}\ \  ,\quad |q|<1\ .\label{ap1}
\end{eqnarray} 
A similar rearrangement to (\ref{trans}) for $|q|>1$ would give
\begin{eqnarray} 
a_+(z) &=& \frac{(1-z^2)}{r}\cdot\prod_{{\rm odd}\ m\ge 1}
\frac{(1-z^2q^{-2(m+1)})^2}{(1-z^2q^{-2m})^2}\ \ ,\quad |q|>1\ .\label{ap2}
\end{eqnarray} 
These are the {\it unique} (up to a sign)  solutions of (\ref{def2})
which are analytic for $z$ in a neighbourhood of
zero in $\Bbb C^*$. In the same way we can analytically extend the series (\ref{equminus})
to obtain $a_-(z)=\pm1/a_+(1/z)$,
 the {\it unique} solutions of (\ref{def}) which are analytic for $z$ in a neighbourhood of
infinity in $\Bbb C^*$.

\end{document}